\documentclass[prl,twocolumn,showpacs,amssymb]{revtex4}

\usepackage{psfrag,graphicx}
\usepackage{tikz}
\usepackage{pgfplots}
\usepackage{dcolumn}
\usepackage{bm}
\usepackage{amsfonts,amssymb,amsmath} 
\usepackage{xcolor}
\usepackage{tikz}
\usetikzlibrary{decorations.markings}
\usetikzlibrary{arrows}
\usepackage{epstopdf}
\usepackage{mathbbol}

\newcommand{\be}{\begin{equation}}
\newcommand{\ee}{\end{equation}}
\newcommand{\bq}{\begin{eqnarray}}
\newcommand{\eq}{\end{eqnarray}}

\newcommand{\no}{\nonumber\\}

\newcommand{\rf}[1]{(\ref{#1})}

\newcommand{\tr}{\mathrm{tr}}
\newcommand{\ket}[1]{\left |#1 \right\rangle}
\newcommand{\bra}[1]{\left \langle #1 \right |}
\newcommand{\braket}[2]{\left \langle #1 \right| \left.#2\right\rangle}

\usetikzlibrary{positioning,shapes.misc}

\tikzset{
    partial ellipse/.style args={#1:#2:#3}{
        insert path={+ (#1:#3) arc (#1:#2:#3)}
    }
}

\begin{document}

\title{Entropic Topological Invariants in Three Dimensions}

\author{Alex Bullivant}
\affiliation{School of Physics and Astronomy, University of Leeds, Leeds, LS2 9JT, United Kingdom}
\author{Jiannis K. Pachos}
\affiliation{School of Physics and Astronomy, University of Leeds, Leeds, LS2 9JT, United Kingdom}

\date{\today}

\pacs{02.40.Pc, 03.65.Vf,65.40.Gr}

\begin{abstract}

\noindent
We evaluate the entanglement entropy of exactly solvable Hamiltonians corresponding to general families of three-dimensional topological models. We show that the modification to the entropic area law due to three-dimensional topological properties is richer than the two-dimensional case. In addition to the reduction of the entropy caused by non-zero vacuum expectation value of contractible loop operators a new topological invariant appears that increases the entropy if the model consists of non-trivially braiding anyons. As a result the three-dimensional topological entanglement entropy provides only partial information about the two entropic topological invariants.

\end{abstract}

\maketitle

%#################################################################################################################################

{\bf \em Introduction:--} The topological features of the entanglement entropy of two-dimensional systems is well understood~\cite{Hamma,KitaevPreskill,LevinWen,Dong}. Tracing a large, smooth region $A$ of a gapped system prepared at its ground state gives the entanglement entropy
\be
S_A = \alpha |\partial A| -b_0 \gamma_0,
\label{eqn:2dim}
\ee
where $\alpha$ and $\gamma_0$ are non-negative constants, $|\partial A|$ is the size of the boundary of $ A$ and the 0-th Betti number, $b_0$, counts the number of disjoint boundary components~\cite{Nakahara}. A non-zero $\gamma_0$ signals that the model is topological in the sense that it supports large loop operators with non-vanishing vacuum expectation value~\cite{LevinWen}. As a consequence it exhibits topological degeneracy and it supports anyonic excitations~\cite{PachosB}. Interestingly, $\gamma_0$ can be isolated by linear combinations of the entanglement entropy corresponding to suitable partitions of the system so that all the boundary terms from the entropy cancel out \cite{KitaevPreskill,LevinWen}. This entropic feature of the ground state, known as the topological entanglement entropy, is one of the main tools used to probe topological order in two dimensions.

The entropic properties of three-dimensional topological models have recently attracted considerable interest~\cite{CastelnovoChamon,Turner,Simon,BenKim}. It has been shown that, even if three dimensions can support string-like topological excitations, their contribution to the topological entanglement entropy is equivalent to the one obtained by point-like excitation~\cite{CastelnovoChamon}. Furthermore, it has been demonstrated that in three dimensions it is possible to have non-universal contributions to the entropy due to a finite correlation length that are constant with the size of the partition~\cite{Turner}. Nevertheless, the constant contribution to the entropy is always linear in $b_0$ and $b_1$, the 0-th and 1-st Betti numbers \cite{Turner}. The latter counts how many topologically inequivalent non-contractible loops can exist on the boundary $\partial A$, e.g. it is zero for the sphere and two for the torus~\cite{Nakahara}. Intriguingly, it has been observed that excitations of some three-dimensional models are confining, significantly affecting the entropic behaviour of the model~\cite{Simon}. Nevertheless, explicit expressions of the entropy have been found for the limited examples of the toric code and the semion model~\cite{Simon} as well as the discrete gauge theories~\cite{Turner}.

Here we evaluate the entanglement entropy of Walker-Wang models that exhibit rich topological behaviours \cite{WalkerWang}. These three-dimensional lattice models are analytically tractable, they have zero correlation length and they are fixed points of general families of three-dimensional topological models. By evaluating explicitly the geometric entanglement entropy we find the general form $S_A = \alpha|\partial A| -b_0\gamma_0 + {b_1\over 2}\gamma_1$, where $\gamma_1$ is a non-negative number. As in two-dimensions the constant $\gamma_0$ determines the total quantum dimensions of the underlying anyonic model of the Walker-Wang model. 
%It is equivalent to indicating if the expectation values of loop operators, corresponding to arbitrarily large but contractible loops, can have a non-zero value. 
It also indicates if anyonic excitations can emerge either in the bulk of the model or at its boundary~\cite{Simon}. If $\gamma_0>0$ then the quantity $\gamma_1$ signals if the anyons of the model braid trivially or not. Both $\gamma_0$ and $\gamma_1$ are {\em entropic topological invariants} that provide information about the topological properties of the system. Surprisingly, we find that topological entanglement entropy does not fully identify the topological properties of the model as it fails to distinguish between the different types of entropic topological invariants.
%While for some Walker-Wang models the topological entanglement entropy behaves in the same way as in two dimensions, for others it vanishes if they support anyons with rich enough braiding statistics. 
%This suggests that the three-dimensional entanglement entropy needs a different approach than its two-dimensional counterpart.  
%Our work is an exposition of the possible entropic behaviours topological systems can have in three dimensions.
Our work provides qualitative and quantitative investigations of the universal topological characteristics of three-dimensional fixed point models.

%#################################################################################################################################

{\bf \em The model:--} Consider an anyonic model with charges $\{1,...,n\}$, where 1 denotes the vacuum. These charges satisfy the fusion rules $a\times b=\sum_cN_{ab}^c c$, where the integers $N_{ab}^c$ denote the multiplicity of the fusion channels. To each anyon $a$ we can assign a real number, the quantum dimension $d_a$, that satisfies $d_ad_b =\sum_cN_{ab}^cd_c$. Moreover, each anyon $a$ has a specific spin, giving rise to a complex phase factor $\theta_a$ when it is rotated counterclockwise around itself by $2\pi$. 
%Rearranging the pairwise order of fusion of four anyons, $a$, $b$, $c$ and $d$, gives rise to the $F$-matrix that relates their in-between fusion outcomes. A single anti-clockwise exchange of two anyons $a$ and $b$, gives rise to the $R$-matrix with diagonal elements enumerated by their different fusion outcomes $c$~\cite{PachosB}. 
%The elements of the $F$ and $R$ matrices are given pictorially by {\bf make it shorter}
The fusion and braiding properties of anyons are described by the $F$- and $R$-matrices, respectively, with elements given pictorially by
\begin{equation}
%\begin{align}
\begin{tikzpicture}[baseline={([yshift=-20pt]current bounding box.north)}]
\draw (0,0) to (0,1);
\draw (0.5,0) to (0.5,1);
\draw (0.5,0.25) to (0,0.75);
\node[font=\fontsize{7}{0}\selectfont] at (-0.15,0) {$c$};
\node[font=\fontsize{7}{0}\selectfont] at (-0.15,1) {$a$};
\node[font=\fontsize{7}{0}\selectfont] at (0.65,0) {$d$};
\node[font=\fontsize{7}{0}\selectfont] at (0.65,1) {$b$};
\node[font=\fontsize{7}{0}\selectfont] at (0.28,0.65) {$e$};
\end{tikzpicture}
\!\!=\sum_f F^{ab,e}_{cd,f}\!
\begin{tikzpicture}[baseline={([yshift=-20pt]current bounding box.north)}]
\draw (0,0) to (0.25,0.25);
\draw (0.5,0) to (0.25,0.25);
\draw (0.25,0.25) to (0.25,0.75);
\draw (0.25,0.75) to (0,1);
\draw (0.25,0.75) to (0.5,1);
\node[font=\fontsize{7}{0}\selectfont] at (-0.15,0) {$c$};
\node[font=\fontsize{7}{0}\selectfont] at (-0.15,1) {$a$};
\node[font=\fontsize{7}{0}\selectfont] at (0.65,0) {$d$};
\node[font=\fontsize{7}{0}\selectfont] at (0.65,1) {$b$};
\node[font=\fontsize{7}{0}\selectfont] at (0.35,0.5) {$f$};
\end{tikzpicture},\,\,
\begin{tikzpicture}[baseline={([yshift=-20pt]current bounding box.north)}]
\draw (0,0) to (0,0.25);
\draw (0,0.25) to (0.25,0.5);
\draw (0.25,0.5) to (0.05,0.7);
\draw (-0.05,0.8) to (-0.25,1);
\draw (0,0.25) to (-0.25,0.5);
\draw (-0.25,0.5) to (0.25,1);
\node[font=\fontsize{7}{0}\selectfont] at (0.15,0.1) {$c$};
\node[font=\fontsize{7}{0}\selectfont] at (0.35,1) {$a$};
\node[font=\fontsize{7}{0}\selectfont] at (-0.35,1.03) {$b$};
\end{tikzpicture}
\!\!=R^{ab}_{c}\!
\begin{tikzpicture}[baseline={([yshift=-20pt]current bounding box.north)}]
\draw (0,0) to (0,0.5);
\draw (0,0.5) to (0.25,1);
\draw (0,0.5) to (-0.25,1);
\node[font=\fontsize{7}{0}\selectfont] at (0.15,0.1) {$c$};
\node[font=\fontsize{7}{0}\selectfont] at (0.35,1.02) {$b$};
\node[font=\fontsize{7}{0}\selectfont] at (-0.35,1) {$a$};
\end{tikzpicture}.
%\end{align}
\end{equation}
Two successive braidings give $R^{ab}_cR^{ba}_c={\theta_c \over \theta_a \theta_b}$ also known as the monodromy. With the monodromy matrix we can introduce the modularity condition. For modular models each anyon $a\neq1$ has a monodromy operator $R^{ab}_cR^{ba}_c$ which is not equal to the identity for at least one charge $b$. In other words a modular model has anyons that braid non-trivially with each other. Otherwise the model is called non-modular. If all pairs of anyons in a model braid trivially with each other then the model is called {\em degenerate} non-modular (or symmetric).

%We consider now the following process. We create from the vacuum two anyon-antianyon pairs $a, \bar a$ and $b, \bar b$. Then we exchange $a$ with $b$ twice before fusing the corresponding pairs together. Each of the latter fusions can result in a new particle, $c$ and $\bar c$ respectively, that subsequently fuse to the vacuum. 
We next define ${\cal S}_{ab}^c$ \cite{AnyonCondensation,KitaevHoney}, in terms of the following quantum mechanical amplitude
\begin{equation}
{\cal S}_{ab}^c = {1 \over \cal D} \!
\begin{tikzpicture}[baseline={([yshift=-15pt]current bounding box.north)}]
\draw [black,domain=80:400] plot ({0.2*cos(\x)}, {0.2*sin(\x)});
\draw [black,domain=-100:220] plot ({0.2+0.2*cos(\x)}, {0.2*sin(\x)});
\draw [black,domain=-5:185] plot ({0.105+0.2*cos(\x)}, {0.2+0.2*sin(\x)});
\node[font=\fontsize{7}{0}\selectfont] at (-0.25,-0.15) (A) {$a$};
\node[font=\fontsize{7}{0}\selectfont] at (0.45,-0.12) (B) {$b$};
\node[font=\fontsize{7}{0}\selectfont] at (0.35,0.35) (C) {$c$};
\end{tikzpicture}
\! = 
\frac{1}{\cal D}\sum_j N^j_{ab}F^{ab,c}_{ab,j}{\theta_j\over \theta_a \theta_b} \sqrt{d_a d_b d_j},
\label{eqn:STensor}
\end{equation}
where ${\cal D} = \sqrt{\sum_a d_a^2}$ is the total quantum dimension of the model. The amplitudes ${\cal S}_{ab}^c$ can be considered as the elements of the {\em ${\cal S}$-tensor}. This tensor is a generalisation of the ${\cal S}$-matrix that has elements ${\cal S}_{ab} \equiv {\cal S}_{ab}^{c=1}$~\cite{KitaevHoney}. The modularity condition is ensured if the ${\cal S}$-matrix is unitary~\cite{Turaev,Bakalov}. This allows us to define the modularity condition
\be
\begin{aligned}
{1 \over {\cal D}^2} \sum_a d_a
\begin{tikzpicture}[baseline={([yshift=-15pt]current bounding box.north)}]
\draw[black] (0,0.1) [partial ellipse=100:440:0.2cm and 0.1cm];
\draw[black] (0,0.05) to (0,0.325);
\draw[black] (0,-0.05) to (0,-0.2);
\node[font=\fontsize{7}{0}\selectfont] at (0.3,0.1) (A) {$a$};
\node[font=\fontsize{7}{0}\selectfont] at (0.085,0.4) (B) {$b$};
\end{tikzpicture}
= \delta_{b1}.
\label{eqn:trap}
\end{aligned}
\ee
In other words the non-trivial braiding of anyon $b$ with some other anyon $a$ causes it to be projected when a symmetrisation is applied.

We now introduce the Walker-Wang models \cite{WalkerWang}. They are three-dimensional generalisations of the string-net models \cite{StringNet}. We consider a three-dimensional trivalent lattice and adopt a certain anyonic model with $n$ different types of charges. We assign an $n$-dimensional Hilbert space at each link, each state corresponding to a Wilson line of a certain anyon charge. It is possible to assign a Hamiltonian to this lattice that gives rise to a specific, possibly degenerate, ground state $\ket{\Phi}$, separated from excited states by a non-zero energy gap. The Hamiltonian energetically penalises the states of the links surrounding a certain vertex that do not satisfy the fusion rules. It also penalises any non-trivial flux that goes through the plaquettes of the lattice \cite{WalkerWang}. The ground state of this Hamiltonian can be considered as a superposition of arbitrary three-dimensional string-net configurations that satisfy the anyonic fusion rules at its vertices. 

Consider the ground state $\ket{\Phi} = \sum_L \Phi(L) \ket{L}$, where $L$ is a certain string-net configuration appearing with amplitude $\Phi(L)$. We can determine these amplitudes in the following way~\cite{StringNet}. Apart from the conditions imposed in the two-dimensional string-net models that implement topological invariance of string configurations, scale invariance and change of basis by employing the $F$-matrices~\cite{StringNet} the three-dimensional Walker-Wang ground states satisfy the additional condition
\be
\begin{aligned}
\Phi (\!\!
\begin{tikzpicture}[baseline={([yshift=-14pt]current bounding box.north)}]
\draw [black,domain=80:400] plot ({0.2*cos(\x)}, {0.2*sin(\x)});
\draw [black,domain=-100:220] plot ({0.2+0.2*cos(\x)}, {0.2*sin(\x)});
\draw [black,domain=-5:185] plot ({0.105+0.2*cos(\x)}, {0.2+0.2*sin(\x)});
\node[font=\fontsize{7}{0}\selectfont] at (-0.25,-0.15) (A) {$a$};
\node[font=\fontsize{7}{0}\selectfont] at (0.45,-0.12) (B) {$b$};
\node[font=\fontsize{7}{0}\selectfont] at (0.35,0.35) (C) {$c$};s
\end{tikzpicture}
\begin{tikzpicture}[every node/.style={fill=lightgray, rectangle, rounded corners, minimum width=0.5cm, minimum height=0.7cm},baseline={([yshift=-12pt]current bounding box.north)}]
    \node[] at (0,0.1)(A){};
\end{tikzpicture}
\,
) ={\cal D} {\cal S}_{ab}^{c}\Phi (\,
\begin{tikzpicture}[every node/.style={fill=lightgray,rectangle,rounded corners, minimum width=0.5cm, minimum height=0.7cm},baseline={([yshift=-12pt]current bounding box.north)}]
    \node[] at (0,0.1) (A){};
\end{tikzpicture}
\,) 
\label{eqn:Sn}
\end{aligned}.
\ee
The configuration $
\begin{tikzpicture}[baseline={([yshift=-15pt]current bounding box.north)}]
\draw [black,domain=80:400] plot ({0.2*cos(\x)}, {0.2*sin(\x)});
\draw [black,domain=-100:220] plot ({0.2+0.2*cos(\x)}, {0.2*sin(\x)});
\draw [black,domain=-5:185] plot ({0.105+0.2*cos(\x)}, {0.2+0.2*sin(\x)});procedure
%\node[font=\fontsize{7}{0}\selectfont] at (-0.25,-0.15) (A) {$a$};
%\node[font=\fontsize{7}{0}\selectfont] at (0.45,-0.12) (B) {$b$};
%\node[font=\fontsize{7}{0}\selectfont] at (0.35,0.35) (C) {$c$};
\end{tikzpicture}
$ is a possible string-net in three dimensions, but not in two dimensions due to its crossings. It has been shown that the behaviour of the point-like excitations of the Walker-Wang models strongly depends on the modularity condition~\cite{Simon}. The modular models have all point-like excitations in the bulk confined as their non-trivial braiding causes a non-zero string tension to emerge. On the other hand, the non-modular models have deconfined bulk excitations. Nevertheless, both models have non-zero entropic topological invariants as we shall see in the following.

%#################################################################################################################################

{\bf \em Geometric entanglement entropy:--} To determine the geometric entanglement entropy we evaluate the von-Neumann entropy $S_A = -\tr (\rho_A \ln\rho_A) $ of the reduced density matrix $\rho_A$ of a geometric region $A$. In two dimensions the entropy $S_A$ is given by \rf{eqn:2dim}, where $\gamma_0 =\ln {\cal D}^2$. Below we show that in three dimensions the entanglement entropy depends on the topology of the boundary $\partial A$ and on the modularity property of the anyonic model.

%#################################################################################################################################

%{\bf Simply connected regions:} 
Consider first tracing a simply connected region $A$ out of the three-dimensional lattice, which is topologically equivalent to a sphere. We take the boundary $\partial A$ to cross the links of the lattice. For convenience we double the Hilbert space of the boundary links so they are present in both $A$ and its compliment $B$. We can write the Schmidt decomposition of the ground state $\ket{\Phi}$ of the system in terms of the states of the corresponding regions
\be
\ket{\Phi} = \sum_{\bm{i}} \alpha_{\bm{i}}\ket{\Phi_A^{\bm{i}}}\ket{\Phi_B^{\bm{i}}},
\label{eqn:schmidt}
\ee
where $\bm{i}$ parametrises the states of all the links of the boundary $\partial A$, $ \sum_{\bm{i}} |\alpha_{\bm{i}}|^2=1$ and $\braket{\Phi_{A/B}^{\bm{i}}}{\Phi_{A/B}^{{\bm{i}'}}} = \delta_{\bm{i},\bm{i}'}$. The entropy of the reduced density matrix $\rho_A = \tr_B(\ket{\Phi}\!\bra{\Phi})$ corresponds to all possible anyonic configurations of the boundary links. As the system is prepared in its ground state these links are subject to the constraint that they all fuse to the vacuum \cite{StringNet}. The probability of having a certain link configuration $\bm{i}$ of anyonic states at the boundary $\partial A$ is given by $P(\bm{i})=|\alpha_{\bm{i}}|^2$. To evaluate $P(\bm{i})$ we first note that the probability of measuring a given charge $a$ at a certain link is given by $P_a=d_a^2/{\cal D}^2$. To impose the fusion constraint at the boundary we introduce the conditional probability of fusing $N$ anyons to the vacuum is given by $P(a_1\times a_2\times ... \times a_{N} \rightarrow 1) = {\cal{N}}_{\bm{a}}^1/(\prod_l d_{a_l})$. Here ${\cal{N}}_{\bm{a}}^c = N_{a_1a_2}^{j_1}N_{j_1a_3}^{j_2}...N_{j_{N-2}a_N}^{c}$ is the total multiplicity associated to the fusion of the $N$ anyons. Therefore the probability that the configuration $\bm{i}$ occurs at the boundary with vacuum total charge is given by $P(\bm{i})  = P_{a_1}...P_{a_{N}} P(a_1\times ... \times a_{N} \rightarrow 1)/P_1$~\cite{suppmat}. In the case where there are several disjoint boundaries comprising $\partial A$ then each boundary component carries vacuum total charge. Hence, the total probability $P(\bm{i})$ is the product of the individual probabilities giving finally
\be
|\alpha_{\bm{i}}|^2 = \frac{{\cal{N}}_{\bm{a}}^1\prod_{l\in {\bm{i}}}d_{a_l}}{ {\cal D}^{2(N-b_0)}}.
\ee
The eigenvalues of the reduced density matrix $\rho_A$ are given by $\lambda_{\bm \sigma} = |\alpha_{\bm{i}}|^2/{\cal{N}}_{\bm{a}}^1$, where ${\bm \sigma} = ({\bm i}, {\bm \mu})$ with ${\bm \mu}$ parametrising the multiplicities of the fusion channels. Hence, the entanglement entropy is given by
\be
S_A = -N\sum_k {d_k^2 \over {\cal D}^2} \ln {d_k \over {\cal D}^2} -b_0 \ln{\cal D}^2.
\label{eqn:simplycon}
\ee
This is the same behaviour as in the two-dimensional case \rf{eqn:2dim}, where $|\partial A| = N$ is the number of links on the boundary of $A$ \cite{LevinWen}. %The $b_0$-term corresponds to the long range topological order present in the ground state, which reduces the overall entanglement entropy.

%#################################################################################################################################

\begin{figure}[t]
\begin{center}
\includegraphics[scale=0.35]{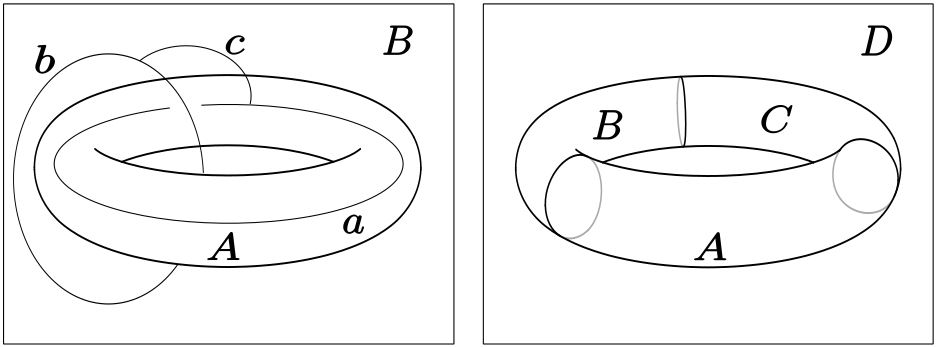}
\end{center}
\caption{\label{fig:Regions} (Left) A toroidal region traced out of the lattice. A possible string-net configuration is depicted with anyon $a$ threading through the inside of the torus and $b$ threads around it. Anyon $c$ is a possible exchanged anyon between the two loops according to the ${\cal S}$-tensor. (Right) Example of a possible configuration of regions $A$, $B$, $C$ and $D$ that extract the topological entanglement entropy in three dimensions.}
\end{figure}

%
%\begin{figure}[t]
%\begin{center}
%\includegraphics[scale=0.38]{Tracing.pdf}
%\end{center}
%fi\caption{\label{fig:Tracing} {\bf Fix Figures} (Left) ... (Right) ...}
%\end{figure}
%{\bf Non-simply connected regions:} 
Consider now the case where the boundary $\partial A$ is topologically equivalent to a torus, as shown in Fig. \ref{fig:Regions} (Left). Compared to the entropic behaviour of simply connected regions the torus can get additional contributions emerging when loops that have support in $A$ and loops that have support in $B$ are braided \cite{Simon}. 
%In this case the additional contribution from the $S$-tensor needs to be considered as we shall see in the following. 
%As these links cannot be removed by unitary operations that have support exclusively in $A$ or in $B$ they contribute non-trivially to the entropy.
To facilitate the calculation of the entropy $S_A$ we bring the state of the system to a suitable form by applications of loop operations that have support exclusively in $A$ or in $B$. These operations do not change the value of $S_A$. We can thus bring the braided loops that belong to different regions in the form shown in Fig. \ref{fig:Regions} (Left). There, $a/b$ is the total anyonic charge of the non-contractible loops that have support in $A/B$, respectively, and $c$ is the exchanged anyon, as dictated by the ${\cal S}$-tensor. The charges of $a$, $b$ and $c$ cannot be changed by non-boundary operations. Thus, according to \eqref{eqn:Sn} the ground state of the system is entropically equivalent to
\be
\ket{\Phi} = {1 \over {\cal D}}\sum_{{\bm{i}}_c,a,b,c} \alpha_{{\bm{i}}_c} {\cal S}_{ab}^c \ket{\Phi_{A,a,c}^{{\bm{i}}_c}}\ket{\Phi_{B,b,c}^{{\bm{i}}_c}},
\label{eqn:psis}
\ee
where ${\bm{i}}_c$ denotes the boundary configuration with a charge $c$ crossing the boundary and $\ket{\Phi_{A,a,c}^{{\bm{i}}_c}}$ and $\ket{\Phi_{B,b,c}^{{\bm{i}}_c}}$ are basis states satisfying $\braket{\Phi_{A/B,a/b,c}^{{\bm{i}}_c}}{\Phi_{A/B,a'/b',c'}^{{\bm{i}}_c'}} = \delta_{{\bm{i}}^{}_c,{\bm{i}}_c'}\delta_{a/b,a'/b'}\delta_{c,c'}$. Note that $\ket{\Phi}$ is properly normalised as 
\be
\sum_{a,b,c} ({\cal S}_{ab}^c)^* {\cal S}_{ab}^c ={\cal D}^2
\label{eqn:id1}
\ee
due to the unitarity of the $F$-matrix~\cite{Bonderson}. We next show that the entanglement entropy of a degenerate non-modular model corresponding to a toroidal boundary is the same as for the simply connected region before we turn to the modular case.

%#################################################################################################################################

%\begin{figure}[t]
%\begin{center}
%\includegraphics[scale=1]{fig1}
%\,\,\,\,\,\,\,\,\,\,\,\,\,\,
%\includegraphics[scale=0.7]{fig2}
%\end{center}
%\caption{\label{fig:Regions} (Left) A toroidal part of the lattice with $T^2$ boundary traced out of the lattice. A possible string-net configuration is presented with anyon $a$ threading %through the inside of the torus and $b$ threads around it. Anyon $c$ is a possible exchanged anyon between the two loops according to the $S$-tensor {\bf make the region A well %distinguished and indicate A and B as this figure is referenced first to show the region. The loops of the anyons are referenced second.} (Right) Example of a possible configuration of regions $A$, $B$, $C$ and $D$ that extract the topological entanglement entropy in three-dimensions. {\bf add region D}}
%\end{figure}

%{\em Degenerate non-modular models:} 
To calculate the entanglement entropy of the degenerate non-modular case we first determine the ${\cal S}$-tensor. As these models have trivial monodromy between all anyons ${\cal S}_{ab}^{c}=\delta_{c1}d_{a}d_{b}/\cal{D}$. Hence, the state \eqref{eqn:psis} takes the form
\begin{equation}
\ket{\Phi} = \frac{1}{{\cal{D}}^2}\sum_{{\bm{i}}_1,a,b}\alpha_{{\bm{i}}_1} d_{a}d_{b}\ket{\Phi_{A,a,1}^{{\bm{i}}_1}}\ket{\Phi_{B,b,1}^{{\bm{i}}_1}}.
\label{eqn:nmd}
\end{equation}
By introducing the orthonormal states $\ket{\tilde\Phi_{A}^{{\bm{i}}_1}}=\sum_{a}\frac{d_{a}}{\cal{D}}\ket{\Phi_{A,a,1}^{{\bm{i}}_1}}$ and $\ket{\tilde \Phi_{B}^{{\bm{i}}_1}} = \sum_{b}\frac{d_{b}}{\cal{D}}\ket{\Phi_{B,b,1}^{{\bm{i}}_1}}$ we can rewrite the ground state \rf{eqn:nmd} in the same form as \eqref{eqn:schmidt}. Hence, for a degenerate non-modular model the entropy of a toroidal region is the same as the entropy of a simply connected region given in \rf{eqn:simplycon}.

%##################################################################################################################################

%{\em Modular models:} 
We now consider the case of modular models. To write the ground state \eqref{eqn:psis} of the modular case in a more convenient form we introduce the following states
$
\ket{\tilde \Phi_{A,bc}^{{\bm{i}}_c}} = {1 \over \sqrt{d_c}} \sum_a {\cal S}_{ab}^c \ket{\Phi_{A,ac}^{{\bm{i}}_c}}. 
$
These states are orthonormal due to the relation
\be
\sum_a({\cal S}_{ab}^c)^* {\cal S}_{ab'}^c = d_c\delta_{bb'},
\label{eqn:id2}
\ee
which can be shown by employing the modularity condition \rf{eqn:trap} \cite{KitaevHoney,suppmat}. Then the ground state of the system becomes 
\be
\ket{\Phi} = \sum_{{\bm{i}}_c,b,c} \alpha_{{\bm{i}}_c}{\sqrt{d_c}\over{\cal D}}\ket{\tilde\Phi_{A,bc}^{{\bm{i}}_c}}\ket{\Phi_{B,bc}^{{\bm{i}}_c}}.
\label{eqn:dec1}
\ee
We can proceed in the evaluation of the entropy in a similar fashion as we did for the simply connected case. Note though that the condition of having the total anyonic charge at the surface $\partial A$ being the vacuum, 1, is now replaced to $c$ due to the braiding of the $a$ and $b$ anyons. The associated probability $P({{\bm{i}}_c}) = |\alpha_{{\bm{i}}_c}|^2$ for the boundary configuration can be written as $P({{\bm{i}}_c}) = P_{a_1}...P_{a_{N}} P(a_1\times ... \times a_{N} \rightarrow c)/P_c$. Combined with the rest of the normalisation factors of state \rf{eqn:dec1} we obtain the eigenvalues of the reduced density matrix $\rho_A$ to be
\be
\lambda_{\bm \sigma} ={ |\alpha_{{\bm{i}}_c}|^2 \over {\cal{N}}_{\bm{a}}^c} {d_c\over {\cal D}^2} = \frac{\prod_{l\in {\bm{i}}_c}d_{a_l}}  {{\cal D}^{2N}}.
\ee
We can directly extend these probabilities to the case of higher genus regions with boundary Betti numbers $b_0$ and $b_1$ by introducing the ${\cal S}$-tensor in \eqref{eqn:psis} multiple times, giving finally the entropy
\be
S_A = - N\sum_k {d_k^2\over {\cal D}^2} \ln {d_k \over {\cal D}^2} - (b_0-{b_1\over 2})\ln{{\cal D} ^2}.
\label{eqn:TorEnt}
\ee
To derive this result we have used the relation $\sum_{b,c} d_c/{\cal D}^2=1$, which follows from \rf{eqn:id1} and \rf{eqn:id2}. For two-dimensional surfaces, such as the boundary $\partial A$, the Betti numbers are related by $\chi = 2b_0-b_1$, where $\chi$ is the Euler characteristic. Hence, the topological dependence of the entropy for modular models is proportional to the Euler characteristic.

%#################################################################

%{\em A family of non-modular models:}
We now consider a family of models with entropic behaviour that ranges between the modular and the degenerate non-modular cases. Take two anyon models, a degenerate non-modular one ${\cal A}_{\rm{DNM}}$ and a modular one ${\cal A}_{\rm{M}}$ with quantum dimensions ${\cal D}_{\rm{DNM}}$ and ${\cal D}_{\rm{M}}$, respectively. We can construct a new anyon model as the direct product ${\cal A} = {\cal A}_{\rm{DNM}} \times {\cal A}_{\rm{M}}$. The charges of such a model are given by $a=\{a_1,a_2\}$, $a_1 \in {\cal A}_{\rm{DNM}}$, $a_2 \in {\cal A}_{\rm{M}}$, with $R^{ab}_{c}=R^{a_{1}b_{1}}_{c_{1}}R^{a_{2}b_{2}}_{c_{2}}$, $F^{ab,e}_{cd,f}=F^{a_{1}b_{1},e_{1}}_{c_{1}d_{1},f_{1}}F^{a_{2}b_{2},e_{2}}_{c_{2}d_{2},f_{2}}$ and total quantum dimension ${\cal D} = {\cal D}_{\rm{DNM}}{\cal D}_{\rm{M}}$. As such ${\cal S}_{ab}^{c}={\cal S}_{a_{1}b_{1}}^{c_{1}}{\cal S}_{a_{2}b_{2}}^{c_{2}}$. 
%Independent of the partition topology the entropy of such models can be written as
%\begin{equation}
%S_A=\sum_{{\bm{i}}_{c_1},{\bm{i}}_{c_2}}P({\bm{i}}_{c_{1}})P({\bm{i}}_{c_{2}})\ln{[P({\bm{i}}_{c_{1}})P({\bm{i}}_{c_{2}})]}.
%\end{equation}
%Here $P({\bm{i}}_{c_{1/2}})$ is the probability amplitude associated with the boundary configuration ${\bm{i}}_{c_{1/2}}$ of anyon model ${\cal A}_{1/2}$. As the two anyon models are non-interacting the boundary probability is normalised for both models independently, $\sum_{{\bm{i}}_{c_{1/2}}}P({\bm{i}}_{c_{1/2}})=1$. 
As the two anyon models are non-interacting the entropy of ${\cal A}$ can be written as the sum of the entropies of each component, $S_A({\cal A})=S_A({{\cal A}_{\rm{DNM}}})+S_A({{\cal A}_{\rm{M}}})$. Then the entanglement entropy is given by
\be
S_A = - N\sum_{k\in {\cal A}} {d_k^2\over {\cal D}^2} \ln {d_k \over {\cal D}^2} - b_0\ln{{\cal D} ^2} + {b_1\over 2} \ln{{\cal D}_{\rm{M}} ^2}.
\label{eqn:general}
\ee
In this way we can construct anyonic models with arbitrary entropic behaviours, ranging between the modular (${\cal D}_{\rm{DNM}} =1$) and the degenerate non-modular (${\cal D}_{\rm{M}}=1$) cases. 
%This generic behaviour is our main result. Below we analyse its properties.

%The linear dependence of the entanglement entropy on the Betti numbers $b_0$ and $b_1$ is a general characteristic \cite{Turner}. 
%For two dimensional surfaces these numbers are related by $\chi = 2b_0-b_1$, where  $\chi$ is the Euler characteristic. 
%With the help of the Gauss-Bonnet theorem we can write the topological number $\chi$ as a surface integral of the Gauss curvature \cite{Nakahara}. 

%#################################################################

{\bf \em Entropic topological invariants:--} 
The entanglement entropy \rf{eqn:general} provides the general form of the entropic topological invariants, $\gamma_0 = \ln {\cal D}^2$ and $\gamma_1 = \ln {\cal D}_{\rm{M}}^2$. The first invariant, $\gamma_0 $, uniquely determines if the model is topological, $\gamma_0\neq 0$, or not, $\gamma_0=0$. Similarly to the two-dimensional case it indicates if loop operators, corresponding to arbitrarily large but contractible loops can have a non-zero expectation value. In accordance to the anyonic fusion rules it reduces the entanglement entropy generated by the superposition of the string-net states. 
%Unlike two dimensions, having non-zero $\gamma_0$ does not mean that the model supports non-trivially braiding anyonic excitations. 
The second invariant, $\gamma_1$, is always less than or equal to $\gamma_0$. It quantifies the total quantum dimensions of anyons in the model which braid non-trivially. If $\gamma_1 =0$ then the model is degenerate non-modular, and all its anyons braid trivially with each other. 
%In three dimensions it is possible to have trivial braiding and non-trivial fusion rules as the hexagon equation need not be imposed. 
These anyons can exist in the bulk of the model as deconfined excitations. If $\gamma_0>\gamma_1$ the model is non-modular supporting both confined and deconfined excitations~\cite{Simon}. If $\gamma_0=\gamma_1>0$ then all excitations of the model are confined and the model is modular. Note that excitations that are confined in the bulk can appear deconfined at the physical boundary of the system~\cite{Simon,SimonPhase}. Topological degeneracy can be manifested by imposing periodic conditions to the three-dimensional bulk or its two-dimensional boundary~\cite{Simon}.

The natural question is what information is gained from the three-dimensional generalisation of the topological entanglement entropy. Similarly to two-dimensional models it is possible to partition the Walker-Wang lattice in four regions, $A$, $B$, $C$ and $D$, such that a combination of their entanglement entropies gives rise to topological invariant quantity. An example of such partition is shown in Fig.~\ref{fig:Regions} (Right). To define a topological invariant of entropies we first introduce the quantity~\cite{Turner}
\begin{align}
\mathfrak{D} [X(A,B,C)] = X_A& + X_B + X_C - X_{AB} - X_{AC} - X_{BC}  \nonumber \\
&\!\!\!\!\!+X_{ABC},
\label{eqn:Dxabc}
\end{align}
where $X$ is some property of the model. For $X=S$ all surface contributions of \rf{eqn:general} cancel out apart from a possible intersection of the four regions~\cite{Turner}. Hence, $\mathfrak{D} [S(A,B,C)]$ is a topological invariant for a choice of regions that do not have such intersections. In this case $\mathfrak{D} [S(A,B,C)]$ is called the topological entanglement entropy $\gamma$. If we now calculate the same quantity for the Euler characteristic, $\chi$, of the boundary we find $\mathfrak{D} [\chi(\partial (A, B, C))] =\chi(A\cap B\cap C \cap D)$ for any choice of four regions~\cite{suppmat}. Demanding topological invariance of $\mathfrak{D} [S(A,B,C)]$ gives $\mathfrak{D} [\chi(\partial (A, B, C))]=0$. From \rf{eqn:general} we obtain the topological entanglement entropy $\gamma = \mathfrak{D} [b_0(\partial(A,B,C))]\ln {\cal D}_{\rm{DNM}}^2$ that can identify the presence of deconfined excitations in the bulk. This analysis also shows that in three dimensions it is not always possible to isolate the topological invariant quantities $\gamma_0$ and $\gamma_1$ by combining the entanglement entropies of a fixed partition. 

{\bf \em Conclusions:--} We have seen that the entanglement entropy of the three-dimensional Walker-Wang models gives two distinct entropic topological invariants, $\gamma_0$ and $\gamma_1$. While $\gamma_0$ identifies if the model is topological, $\gamma_1$ reveals information about the braiding properties of its underlying anyonic model. The latter corresponds to an increase in the entropy as tracing the region around a torus erases information about the anyonic charge of the threaded loops that could affect non-trivially the state of the system inside the torus. Hence, the geometric entanglement entropy in three dimensions provides more information about the topological order of the system than its two-dimensional counterpart.
%Such a term is possible to emerge as the ground state of these three-dimensional models is a superposition of possibly braided string-net configurations, which is not the case in two dimensions. 
We can isolate $\gamma_0$ by comparing the entropy of two different partitions, with different number of disjoint boundary components, $b_0$, but the same $b_1$ and $|\partial A|$. Similarly we can determine $\gamma_1$. Evaluating the entanglement entropy for the general non-modular case is a complex problem due to the lack of structure of these models. We leave this problem to future investigation. %Finally, it is of interest to consider the reliable identification of entropic topological invariants in systems with non-zero correlation length~\cite{Turner}. 

{\bf \em Acknowledgements:--} We would like to thank Gavin Brennen, Ben Brown, Curt von Keyserlingk, Paul Martin, Steven Simon and Jamie Vicary for inspiring conversations. This work was supported by EPSRC.

\bibliographystyle{apsrev4-1}
\bibliography{paper}

%#################################################################
\newpage
%#################################################################

\noindent
{\bf \em Supplementary Material}\\

%#################################################################

%#################################################################

\noindent
{\bf \em A. Probability of boundary configurations:--}
There are two equivalent ways to assign a probability to a given boundary configuration on a region $A$ of a string-net configuration in Walker-Wang models. The first is to use unitary operations with support on either $A$ or the compliment $B$. One can then form a canonical configuration with string-nets restricted to the plaquettes crossing the boundary $\partial A$. Then the diagram calculus can be utilised to calculate the amplitude of each configuration and hence the probability \cite{LevinWen}. The second approach is to assign a conditional probability to the anyon configuration making use of the quantum dimensions of the model \cite{KitaevPreskill}. This is the methodology we utilise here as it is computationally easier .

In order to calculate the probability of finding an anyon charge we make use of the ${\cal S}$-matrix. The probability of finding an anyon of charge $a$ in a particular region is given by $P_a=|{\cal S}_{a1}|^2=d^{2}_{a}/D^2$ \cite{KitaevPreskill}. This result can be understood as the probability amplitude for the vacuum braiding trivially with the charge $a$. Due to charge conservation across the boundary the probability of each anyon on the boundary is not independent but we require that all anyons fuse to a total charge $c$. For a boundary consisting of $N$ charges this constraint is applied by defining the conditional probability $P(a_1\times...\times a_{N}\rightarrow c)$. This probability can be evaluated diagrammatically making use of isotopy invariance \cite{Preskill} and the identity $F^{ab,1}_{ab,c}=\sqrt{\frac{d_c}{d_ad_b}}$~\cite{Bonderson}
\begin{align}
P(a_1a_2\rightarrow j_1)&=P(a_1a_2\rightarrow j_1)\frac{1}{d_{a_1}d_{a_2}}
\begin{tikzpicture}[baseline={([yshift=-17pt]current bounding box.north)}]
\draw (0,0) circle (0.25cm);
\draw (0,0) circle (0.5cm);
\node[font=\fontsize{7}{0}\selectfont] at (-0.05,0) {$a_1$};
\node[font=\fontsize{7}{0}\selectfont] at (-0.65,0.0) {$a_2$};
\end{tikzpicture}
\nonumber\\
&=\frac{N^{j_1}_{a_1a_2}}{d_{a_1}d_{a_2}}\sqrt{\frac{d_{j_1}}{d_{a_1}d_{a_2}}}
\begin{tikzpicture}[baseline={([yshift=-28pt]current bounding box.north)}]
\draw [black,domain=0:180] plot ({0.5*cos(\x)}, {0.75+0.5*sin(\x)});
\draw [black,domain=180:360] plot ({0.5*cos(\x)}, {0.5*sin(\x)});
\draw [black,domain=0:180] plot ({-0.5+0.25*cos(\x)}, {0.5+0.25*sin(\x)});
\draw [black,domain=180:360] plot ({-0.5+0.25*cos(\x)}, {0.25+0.25*sin(\x)});
\draw (-0.75,0.25) to (-0.75,0.5);
\draw (-0.25,0.25) to (-0.25,0.5);
\draw (0.5,0) to (0.5,0.75);
\node[font=\fontsize{7}{0}\selectfont] at (-0.05,0.39) {$a_1$};
\node[font=\fontsize{7}{0}\selectfont] at (-0.9,0.39) {$a_2$};
\node[font=\fontsize{7}{0}\selectfont] at (0.7,0.39) {$j_1$};
\end{tikzpicture}
\nonumber\\
&=\frac{N^{j_1}_{a_1a_2}d_{j_1}}{d_{a_1}d_{a_2}}.\tag{A1}
\label{eqn:pabc}
\end{align}
We now calculate $P(a_1\times...\times a_{N}\rightarrow c)$ as below
\begin{align}
&P(a_1\times...\times a_{N}\rightarrow c)
\nonumber\\
&=P(a_1a_2\rightarrow j_1)P(a_3j_1\rightarrow j_2)...P(a_Nj_{N-2}\rightarrow c)
\nonumber\\
&=\frac{{\cal{N}}_{\bm{a}}^cd_c}{\prod_{l=1}^{N}d_{a_l}}\tag{A2},
\end{align}
where ${\cal{N}}_{\bm{a}}^c=N^{j_1}_{a_1a_2}N^{j_2}_{a_3j_1}...N^{c}_{a_Nj_{N-2}}$.
One can verify the consistency equation
\begin{equation}\tag{A3}
\sum_{a_1,...,a_N\\j_1,...,j_{N-2}}P_a...P_bP(a_1\times...\times p_{N}\rightarrow c)=P_c.
\end{equation}
The normalised probability for the boundary configuration with a given charge $c$ across the boundary is given by
\begin{equation}\tag{A4}
P({\bm{i}}_c)=\frac{P_a...P_bP(a_1\times...\times p_{N}\rightarrow c)}{P_c}.
\end{equation}

{\bf \em B. ${\cal S}$-tensor Properties:--}
The ${\cal S}$-matrix has been extensively studied throughout the literature \cite{Turaev,Bakalov}. Here we outline the generalisation to the ${\cal S}$-tensor and we demonstrate some of its properties. The ${\cal S}$-tensor can be defined diagrammatically or in terms of the parameters of the model as follows \cite{AnyonCondensation}
 \begin{equation}\tag{B1}
 {\cal S}_{ab}^c \equiv {1 \over \cal D} \!
 \begin{tikzpicture}[baseline={([yshift=-15pt]current bounding box.north)}]
 \draw [black,domain=80:400] plot ({0.2*cos(\x)}, {0.2*sin(\x)});
 \draw [black,domain=-100:220] plot ({0.2+0.2*cos(\x)}, {0.2*sin(\x)});
 \draw [black,domain=-5:185] plot ({0.105+0.2*cos(\x)}, {0.2+0.2*sin(\x)});
 \node[font=\fontsize{7}{0}\selectfont] at (-0.25,-0.15) (A) {$a$};
 \node[font=\fontsize{7}{0}\selectfont] at (0.45,-0.12) (B) {$b$};
 \node[font=\fontsize{7}{0}\selectfont] at (0.35,0.35) (C) {$c$};
 \end{tikzpicture}
 \! = 
 \frac{1}{\cal D}\sum_j N^j_{ab}F^{ab,c}_{ab,j}{\theta_j\over \theta_a \theta_b} \sqrt{d_a d_b d_j}.
 \label{eqn:Sabc}
 \end{equation}
One useful property of the ${\cal S}$-tensor is the following statement
\begin{equation}\tag{B2}
D^2=\sum\limits_{a,b,c}({\cal S}^{c}_{ab})^* {\cal S}^{c}_{ab}.
\end{equation}
This follows from Eqn. \eqref{eqn:Sabc} noting the unitarity of the $F$-matrices $\sum_c (F^{ab,c}_{ab,j})^*F^{ab,c}_{ab,j'}=\delta_{j,j'}$~\cite{Bonderson}. In the remainder of the section we make use of the diagram calculus in order to derive properties of the ${\cal S}$-tensor.

We begin our discussion by defining the fusion Hilbert space and the relevant normalisation scheme~\cite{KitaevHoney,Bonderson,Preskill}. For the fusion process $a\times b\rightarrow c$ we can define a fusion Hilbert space $V^{ab}_{c}$ with dimension given by the fusion multiplicities $N^{c}_{ab}=$dim$V^{ab}_{c}$. We define an orthonormal basis for such a space with states $\ket{a,b;c,\mu}\in V^{ab}_{c}$ satisfying $\braket{a',b';c',\mu'}{a,b;c,\mu}=\delta_{aa'}\delta_{bb'}\delta_{cc'}\delta_{\mu\mu'}$. We can represent these states and their duals in terms of the diagram calculus
\begin{align}
\ket{a,b;c,\mu}=
\begin{tikzpicture}[baseline={([yshift=-18pt]current bounding box.north)}]
\draw (0,0) to (-0.25,0.25);
\draw (0.25,0.25) to (0,0);
\draw (0,0) to (0,-0.25);
\node[font=\fontsize{7}{0}\selectfont] at (-0.35,0.35) {$a$};
\node[font=\fontsize{7}{0}\selectfont] at (0.35,0.35) {$b$};
\node[font=\fontsize{7}{0}\selectfont] at (0,-0.35) {$c$};
\node[font=\fontsize{7}{0}\selectfont] at (0.15,-0.1) {$\mu$};
\end{tikzpicture},\,\,
%\no
\bra{a,b;c,\mu}=
\begin{tikzpicture}[baseline={([yshift=-18pt]current bounding box.north)}]
\draw (-0.25,-0.25) to (0,0);
\draw (0.25,-0.25) to (0,0);
\draw (0,0) to (0,0.25);
\node[font=\fontsize{7}{0}\selectfont] at (-0.35,-0.35) {$a$};
\node[font=\fontsize{7}{0}\selectfont] at (0.35,-0.35) {$b$};
\node[font=\fontsize{7}{0}\selectfont] at (0,0.35) {$c$};
\node[font=\fontsize{7}{0}\selectfont] at (0.15,0.05) {$\mu$};
\end{tikzpicture}.\tag{B3}
\end{align}
The inner product is defined diagrammatically by stacking vertices vertically and connecting open edges. To preserve isotopy invariance and normalise the inner products we introduce a constant depending on the quantum dimensions of the edge charges
\begin{align}
\braket{a',b';c',\mu'}{a,b;c,\mu}&=\frac{\delta_{aa'}\delta_{bb'}\delta_{cc'}\delta_{\mu\mu'}}{\sqrt{d_ad_bd_c}}
\begin{tikzpicture}[baseline={([yshift=-27pt]current bounding box.north)}]
\draw [black,domain=0:180] plot ({0.5*cos(\x)}, {0.75+0.5*sin(\x)});
\draw [black,domain=180:360] plot ({0.5*cos(\x)}, {0.5*sin(\x)});
\draw [black,domain=0:180] plot ({-0.5+0.25*cos(\x)}, {0.5+0.25*sin(\x)});
\draw [black,domain=180:360] plot ({-0.5+0.25*cos(\x)}, {0.25+0.25*sin(\x)});
\draw (-0.75,0.25) to (-0.75,0.5);
\draw (-0.25,0.25) to (-0.25,0.5);
\draw (0.5,0) to (0.5,0.75);
\node[font=\fontsize{7}{0}\selectfont] at (-0.05,0.39) {$b$};
\node[font=\fontsize{7}{0}\selectfont] at (-0.9,0.39) {$a$};
\node[font=\fontsize{7}{0}\selectfont] at (0.7,0.39) {$c$};
\node[font=\fontsize{7}{0}\selectfont] at (-0.6,-0.15) {$\mu$};
\node[font=\fontsize{7}{0}\selectfont] at (-0.6,0.85) {$\mu$};
\end{tikzpicture}
\no&=\delta_{aa'}\delta_{bb'}\delta_{cc'}\delta_{\mu\mu'}.\tag{B4}
\end{align}
We also introduce the identity operator
\begin{align}
{\mathbb{1}}=\sum\limits_{a,b,c,\mu}\ket{a,b;c,\mu}\bra{a,b;c,\mu}=\sum\limits_{a,b,c,\mu}\sqrt{\frac{d_c}{d_ad_b}}
\begin{tikzpicture}[baseline={([yshift=-27pt]current bounding box.north)}]
\draw (0,0) to (-0.25,-0.25);
\draw (0,0) to (0.25,-0.25);
\draw (0,0.5) to (0.25,0.75);
\draw (0,0.5) to (-0.25,0.75);
\draw (0,0) to (0,0.5);
\node[font=\fontsize{7}{0}\selectfont] at (-0.35,-0.35) {$a$};
\node[font=\fontsize{7}{0}\selectfont] at (0.35,-0.35) {$b$};
\node[font=\fontsize{7}{0}\selectfont] at (-0.15,0.275) {$c$};
\node[font=\fontsize{7}{0}\selectfont] at (0.15,0.05) {$\mu$};
\node[font=\fontsize{7}{0}\selectfont] at (-0.35,0.85) {$a$};
\node[font=\fontsize{7}{0}\selectfont] at (0.35,0.85) {$b$};
\node[font=\fontsize{7}{0}\selectfont] at (0.15,0.4) {$\mu$};
\end{tikzpicture}.\tag{B5}
\end{align}
We now represent the ${\cal S}$-tensor as an operator ${\cal S}_c$ on the fusion space $V^{bc}_{c}$ as defined in \cite{KitaevHoney}
\begin{align}
{\cal S}_c\ket{b,c;b,\mu}=\sum\limits_{x}\frac{d_x}{\cal{D}}
\begin{tikzpicture}[baseline={([yshift=-22pt]current bounding box.north)}]
\draw [black,domain=-90:250] plot ({0.25*cos(\x)}, {0.25*sin(\x)});
\draw [black,domain=-35:15] plot ({-0.75+0.75*cos(\x)}, {0.75*sin(\x)});
\draw [black,domain=23:40] plot ({-0.75+0.75*cos(\x)}, {0.75*sin(\x)});
\draw (0.25,0) to (0.5,0.25);
\node[font=\fontsize{7}{0}\selectfont] at (0.6,0.35) {$c$};
\node[font=\fontsize{7}{0}\selectfont] at (0.35,-0.1) {$\mu$};
\node[font=\fontsize{7}{0}\selectfont] at (0.25,0.3) {$b$};
\node[font=\fontsize{7}{0}\selectfont] at (0.25,-0.3) {$b$};
\node[font=\fontsize{7}{0}\selectfont] at (-0.25,-0.5) {$x$};
\node[font=\fontsize{7}{0}\selectfont] at (-0.3,0.55) {$x$};
\end{tikzpicture}.\tag{B6}
\end{align}
Utilising the diagram calculus and normalisation conventions defined above we can form the matrix elements of the ${\cal S}$-tensor operator
\begin{equation}\tag{B7}
{\cal S}_{ab}^{c}=\sqrt{d_c}\bra{a,c;a,\mu}{\cal S}_c\ket{b,c;b,\mu}.
\label{eqn:sbraket}
\end{equation}

The above discussion of the ${\cal S}$-tensor holds true for any anyon model. In the following we restrict our discussion to modular anyon models to show that for such models
\begin{equation}\tag{B8}
\sum\limits_{a}({\cal S}_{ab'}^{c})^* {\cal S}_{ab}^{c}=d_c\delta_{b,b'}.
\label{eqn:dcmod}
\end{equation}
As discussed in the main text an anyon model is described as being modular when for each non-vacuum charge $a$ in the model there exists a charge $b$ such that the monodromy $R^{c}_{ba}R^{c}_{ab}\neq \mathbb{1}_{ab}$. This statement can be equivalently formulated in the diagram calculus as the modular trap identity \cite{KitaevHoney,Turaev}
\begin{align}
{1 \over {\cal D}^2} \sum_a d_a
\begin{tikzpicture}[baseline={([yshift=-15pt]current bounding box.north)}]
\draw[black] (0,0.25) [partial ellipse=-240:60:0.3cm and 0.15cm];
\draw[black] (0,0.25) [partial ellipse=80:100:0.3cm and 0.15cm];
\draw (-0.1,0.2) to (-0.1,0.5);
\draw (0.1,0.2) to (0.1,0.5);
\draw (-0.1,-0.1) to (-0.1,0.05);
\draw (0.1,-0.1) to (0.1,0.05);
\node[font=\fontsize{7}{0}\selectfont] at (0.4,0.35) {$a$};
\node[font=\fontsize{7}{0}\selectfont] at (0.1,0.62) {$x$};
\node[font=\fontsize{7}{0}\selectfont] at (-0.1,0.6) {$y$};
\end{tikzpicture}
=\frac{\delta_{xy}}{d_x}
\begin{tikzpicture}[baseline={([yshift=-17pt]current bounding box.north)}]
\draw [black,domain=0:180] plot ({0.2*cos(\x)}, {0.2*sin(\x)});
\draw [black,domain=180:360] plot ({0.2*cos(\x)}, {0.6+0.2*sin(\x)});
\node[font=\fontsize{7}{0}\selectfont] at (-0.2,0.7) {$x$};
\node[font=\fontsize{7}{0}\selectfont] at (0.2,0.7) {$x$};
\node[font=\fontsize{7}{0}\selectfont] at (-0.2,-0.1) {$x$};
\node[font=\fontsize{7}{0}\selectfont] at (0.2,-0.1) {$x$};
\end{tikzpicture}.
\label{eqn:trap2}\tag{B9}
\end{align}
Using this identity it is possible to show that the operator ${\cal S}_c$ is unitary for modular models
\begin{align}
{\cal S}_c^\dagger {\cal S}_c\ket{b,c;b,\mu}&=\sum\limits_{x,y}\frac{d_xd_y}{{\cal{D}}^2}
\begin{tikzpicture}[baseline={([yshift=-23pt]current bounding box.north)}]
\draw [black,domain=-115:220] plot ({0.25*cos(\x)}, {0.25*sin(\x)});
\draw [black,domain=65:100] plot ({-0.3+0.25*cos(\x)}, {0.25*sin(\x)});
\draw [black,domain=40:-240] plot ({-0.3+0.25*cos(\x)}, {0.25*sin(\x)});
\draw [black,domain=-15:40] plot ({-1.1+0.75*cos(\x)}, {0.75*sin(\x)});
\draw [black,domain=-22:-35] plot ({-1.1+0.75*cos(\x)}, {0.75*sin(\x)});
\draw (0.25,0) to (0.5,0.25);
\node[font=\fontsize{7}{0}\selectfont] at (0.6,0.35) {$c$};
\node[font=\fontsize{7}{0}\selectfont] at (0.35,-0.1) {$\mu$};
\node[font=\fontsize{7}{0}\selectfont] at (0.25,0.3) {$b$};
\node[font=\fontsize{7}{0}\selectfont] at (0.25,-0.3) {$b$};
\node[font=\fontsize{7}{0}\selectfont] at (-0.55,-0.55) {$y$};
\node[font=\fontsize{7}{0}\selectfont] at (-0.6,0.55) {$y$};
\node[font=\fontsize{7}{0}\selectfont] at (-0.25,0.35) {$x$};
\end{tikzpicture}
\no
&=
\begin{tikzpicture}[baseline={([yshift=-17pt]current bounding box.north)}]
\draw (0,0) to (-0.25,0.25);
\draw (0.25,0.25) to (0,0);
\draw (0,0) to (0,-0.25);
\node[font=\fontsize{7}{0}\selectfont] at (-0.35,0.35) {$b$};
\node[font=\fontsize{7}{0}\selectfont] at (0.35,0.35) {$c$};
\node[font=\fontsize{7}{0}\selectfont] at (0,-0.35) {$b$};
\node[font=\fontsize{7}{0}\selectfont] at (0.15,-0.1) {$\mu$};
\end{tikzpicture}
=\ket{b,c;b,\mu}.\tag{B10}
\end{align}
We then prove \eqref{eqn:dcmod} by invoking the unitary property of ${\cal S}_c$ and \eqref{eqn:sbraket}
\begin{align}
&d_c\delta_{bb'}
\no
=&d_c\braket{b',c;b',\mu}{b,c;b,\mu}
\no
=&d_c\bra{b',c;b',\mu}{\cal S}_c^\dagger {\cal S}_c\ket{b,c;b,\mu}
\no
=&\sum\limits_{a,c',\mu'}d_c\bra{b',c;b',\mu}{\cal S}_c^\dagger\ket{a,c';a,\mu'}\bra{a,c';a,\mu'}{\cal S}_c\ket{b,c;b,\mu}
\no
=&\sum\limits_{a}d_c\bra{b',c;b',\mu}{\cal S}_c^\dagger\ket{a,c;a,\mu}\bra{a,c;a,\mu}{\cal S}_c\ket{b,c;b,\mu}
\no
=&\sum\limits_{a}({\cal S}^{c}_{ab'})^*{\cal S}^{c}_{ab}.\tag{B11}
\end{align}

%#################################################################

\noindent
{\bf \em C. Isolating topological components of the entanglement entropy:--} As for two-dimensional models \cite{KitaevPreskill,LevinWen}, it is possible to partition the Walker-Wang lattice into four regions $A,B,C$ and $D$ from which a combination of entropies can be evaluated to formulate a topologically invariant quantity~\cite{Turner,CastelnovoChamon}. To define a topologically invariant quantity we introduce the quantity $\mathfrak{D} [X(A,B,C)]$ as in \cite{Turner} 
\begin{align}
\mathfrak{D} [X(A,B,C)] = X_A& + X_B + X_C - X_{AB} - X_{AC} - X_{BC}  \no
&\!\!\!\!\!+X_{ABC},
\label{eqn:Dabcap}\tag{C1}
\end{align}
where $X$ is a property of the regions $A,B,C$ and $D$. Work by Grover {\em et al.} \cite{Turner} conjectured that for $X=S$ all surface contributions from the geometric entanglement entropies cancel when the quantity $\mathfrak{D} [\chi(\partial (A, B, C))]=0$, where $\chi$ is the Euler characteristic. In the following we show that under minimal constraints, surface contributions and terms proportional to the Euler characteristic vanish identically in equation \eqref{eqn:Dabcap}. Nevertheless, our workings show that the vanishing of the Euler characteristic does not imply the vanishing of surface contributions as claimed in \cite{Turner}.

Consider the lattice in the continuum limit as a closed subset of $\mathbb{R}^3$ partitioned into four finite regions $A$, $B$, $C$ and $D$ with non-intersecting volumes where $A$, $B$, $C$ are embedded within $D$. We define the condition of non-intersecting volumes by the statement, two volumes $I$ and $J$ are non-intersecting if and only if $I\cap J=\partial I \cap \partial J$ where $\partial I$ is the boundary of $I$. By convention we adopt the following notation for composite regions $IJ=I\cup J$. We utilise the above construction to define the boundary of the regions $A$, $B$, $C$ and their composites with respect to union and intersections of the other regions

\begin{align}
\partial (ABC)&=(ABC)\cap D, \nonumber \\
\partial (AB)&=(AB)\cap(CD), \nonumber \\
\partial A&=A\cap(BCD). \tag{C2}
\end{align}
The boundaries of the other regions can be formulated likewise.

Measures of the surface area and Euler characteristic both obey the so called inclusion-exclusion principle. A property of a set $X$ obeys the inclusion-exclusion principle when $X(AB)=X(A)+X(B)-X(A\cap B)$. Making use of this property one can easily verify
\begin{equation}\tag{C3}
\mathfrak{D} [\chi(\partial (A, B, C))]=X(A\cap B\cap C \cap D).
\end{equation}
When $X$ is taken to be the surface area or Euler characteristic, $Area(\emptyset)=\chi(\emptyset)=0$, hence for $A\cap B\cap C \cap D=\emptyset$, $\mathfrak{D} [X(\partial (A, B, C))]=0$. This result demonstrates that such a combination of boundary areas or Euler characteristic will always vanish for the above construction when there is no simultaneous intersection of all four regions. Furthermore such a quantity is necessarily invariant under smooth deformations of the regions complimentary to the discussion in \cite{Turner}.

\end{document}